\documentclass[twocolumn,showpacs,preprintnumbers,amsmath,amssymb]{revtex4}

\usepackage{graphicx}
\usepackage{dcolumn}
\usepackage{bm}

\begin{document}

\preprint{APS/123-QED}

\title{Mode entanglement of an electron in one-dimensional
determined and random potentials}
\author{Longyan Gong  $^{1,}$  and Peiqing Tong  $^{2,}$ }
\email{pqtong@pine.njnu.edu.cn}
\affiliation{%
 $^{1}$National Laboratory of Solid State Microstructures and
 Department of Physics, Nanjing University, Nanjing, Jiangsu,
 P.R.China\\
 $^{2}$Department of Physics, Nanjing Normal University,
Nanjing, Jiangsu, P.R.China
}%
\date{today}
\begin{abstract}
By using the measure of concurrence, mode entanglement of an
electron moving in four kinds of one-dimensional determined and
random potentials is studied numerically. The extended and
localized states can be distinguished by mode entanglement. There
are sharp transitions in concurrence at mobility edges. It
provides that the mode entanglement may be a new index for a
metal-insulator transition.
\end{abstract}
\maketitle

\section{Introduction}
Entanglement is a unique feature of quantum systems that play a
key role in quantum information processing. The early study of
entanglement is only focused on the foundations of the quantum
mechanics \cite{sch35}. Recently due to its potential applications
in quantum communications, quantum cryptography, quantum computer
and quantum information \cite{bo00}, entanglement has been studied
extensively. One of the most important progress is the
quantitative measures of entanglement for mixed state by using the
entanglement of formation \cite{be96a,be96b}. For the special case
of two spin-1/2 systems, the entanglement of formation is given by
the concurrence \cite{wo01,hi97}. Newly, considerable interest has
been devoted to entanglement of quantum spin system
\cite{vi03,gl03}, identical particles \cite{sc01,wi03}, fractional
quantum Hall effect \cite{ze02}, and spins of a noninteracting
electron gas \cite{sa04}.

On the other hand, since Anderson published his famous paper
\cite{an58} about disorder induced localization, extensive
investigations have focused on the metal-insulator transition
(MIT). For one-dimensional (1D) Anderson model, it is well known
\cite{ra85} that all eigenstates are localized and there is no
mobility edge separating localized and extended states. However,
the specific extended states and/or mobility edges have been found
in several 1D determined and random model with short-range and
long-range correlation \cite{gr88,th88,sa88,du90,mo98,ro03,xi03}.
The well-studied examples of the potentials are a slowly varying
potential \cite{gr88,th88,sa88}, random-dimer potential
\cite{du90}, long-range correlated disordered potential
\cite{mo98} and Anderson model with long-range hopping
\cite{ro03,xi03}. In these models, electronic localized behaviors
are studied by judging Thouless exponent (or Lyapunov
coefficient), participation ratio or dynamics of wave function.

Recently mode entanglement of spinless electrons sharing in
one-particle states in 1D model has been investigated in Refs.
\cite{la03,wa04}. Using the ordinary Harper and the kicked Harper
model, Lakshminarayan and Subrahmanyam found that entanglement can
reflect MIT. Similar behavior is also found for the ground state
of an electron in 1D Frenkel-Kontorova (FK) potential \cite{wa04}.
There are many more complex 1D potentials, e.g., potentials used
in Refs.\cite{gr88,th88,sa88,du90,mo98,ro03,xi03}, which have been
widely used to study the MIT and exhibit more complex localization
behaviors than that of the Harper model. Therefore, it is
interesting to study the mode entanglement of an electron in these
more complex 1D potentials. 

The paper is organized as follows. In next section the formalism
of entanglement and concurrence is described. In Sec.~\ref{sec3}
the numerical results for four kinds of models are presented.
Section ~\ref{sec4} is devoted to conclusion.

\section{\label{sec2}Formalism}
In the second-quantized picture, the Hamiltonian for electrons
moving in 1D determined and/or random potential can be written as
follow:
\begin{equation}
H =  - t\sum\limits_{n = 1}^N {(c_n^ +  c_{n + 1}  + c_{n + 1}^ +
c_n ) + \sum\limits_{n = 1}^N {V_n c_n^ +  c_n } },
\end{equation} 
where $t$ is a nearest-neighbor hopping integral, $c_n^+$ ($c_n$)
is the creation(annihilation) operator of \emph{n}th site, and
$V_n$ is the one-site potential. In our numerical studies, we take
$t=1$ without loss of generality. The site occupation basis is
\begin{equation}
\left| n_1, n_2, \ldots ,n_N
\right\rangle=c_1^{+n_1}c_2^{+n_2}\ldots c_N^{+n_N}\left|
0\right\rangle,
\end{equation}
where $n_i=0, 1$, and $\left|0\right\rangle$ is the vacuum. Note
that there is an isomorphism between these states and the states
of N qubits\cite{la03}. For an electron, $\sum_{i=1}^{N}n_i=1$. If
we write $\left|n \right\rangle=\left| 0,\ldots , 1_n, \ldots ,0
\right\rangle,$
 the general state of an electron is
\begin{equation}
\left| \Psi  \right\rangle  = \sum\limits_{n = 1}^N {\Psi _n }
\left| n \right\rangle  = \sum\limits_{n = 1}^N {\Psi _n c_n^ + }
\left| 0 \right\rangle,
\end{equation} 
where ${\Psi _n }$ is amplitude of wave at \emph{n}th site.

From Eqs. (1), (2) and (3), we obtain the eigenequation
  \begin{equation}
 - (\Psi _{n + 1}  + \Psi _{n - 1})t  + V_n \Psi _n  = E\Psi _n,
  \end{equation} 
where $E$ is the eigenenergy. For an eigenstate $\beta$ with
eigenenergy $E_\beta$, the concurrence between sites ( or qubits)
$i$ and $j$ is given \cite{la03} as
\begin{equation}
C_{ij}^\beta   = 2\left| {\Psi _i^\beta  \Psi _j^\beta  } \right|.
\end{equation} 
States that have a large minimum pairwise concurrence can be said
share entanglement better. Specifically, when $\left|{\Psi
_i^\beta}\right|=1/\sqrt{N}$, the state becomes the so-called $W$
state \cite{du00} and the concurrence is given by $2/N$. There can
not be states whose minimum pairwise concurrence exceeds $2/N$. As
a gross but useful measure of entanglement sharing, Lakshminarayan
and Subrahmanyam propose and study the pairwise concurrence
averagely in a given state. For the given eigenstate,
\begin{equation}
\left\langle {C^\beta  } \right\rangle  =
\frac{1}{d}\sum\limits_{i < j} {C_{ij}^\beta   = }
\frac{1}{d}((\sum\limits_{i = 1}^N {\left| {\Psi _i^\beta  }
\right|} )^2  - 1),
\end{equation} 
where $d=N(N-1)/2$. From the definition of (6), we can see that
$<C^\beta>$ has connections to measures of localization of the
eigenstates. As a further gross measure they also average over all
the eigenstates $\beta$ , i.e. ,
\begin{equation}
\left\langle {C} \right\rangle  = \frac{1}{M}\sum\limits_{\beta}
\left\langle {C^\beta }\right\rangle,
\end{equation} 
where $M$ is the number of all the states. In the following
concurrence $ {\left\langle {C} \right\rangle }$ and
${\left\langle {C^\beta} \right\rangle }$ are measured for four
kinds of 1D determined and random potentials .

\section{\label{sec3}Numerical Results }

\subsection{Slowly varying potential} 

In the slowly varying potential model \cite{gr88,th88,sa88}, the
on-site potential is given by $V_n = \lambda \cos (\pi \alpha
n^\upsilon ) $, here $\lambda$, $\alpha$ and $ 0 \leq \upsilon
\leq1 $ are positive numbers which completely define the
tight-binding problem. If $\upsilon=1$, the model is well known
Harper model. For Harper model, the eigenstates are either all
extended or  all localized states depending on whether $\lambda$
is smaller or larger than 2.0 \cite{sa88}. When $\lambda=2.0$, all
the states are critical. The mode entanglement of Harper model has
been investigated in Ref. \cite{la03} and it was found a sharp
transition in concurrence at $\lambda=2.0$, which corresponds to
MIT.

\begin{figure}
\includegraphics[width=2.5in]{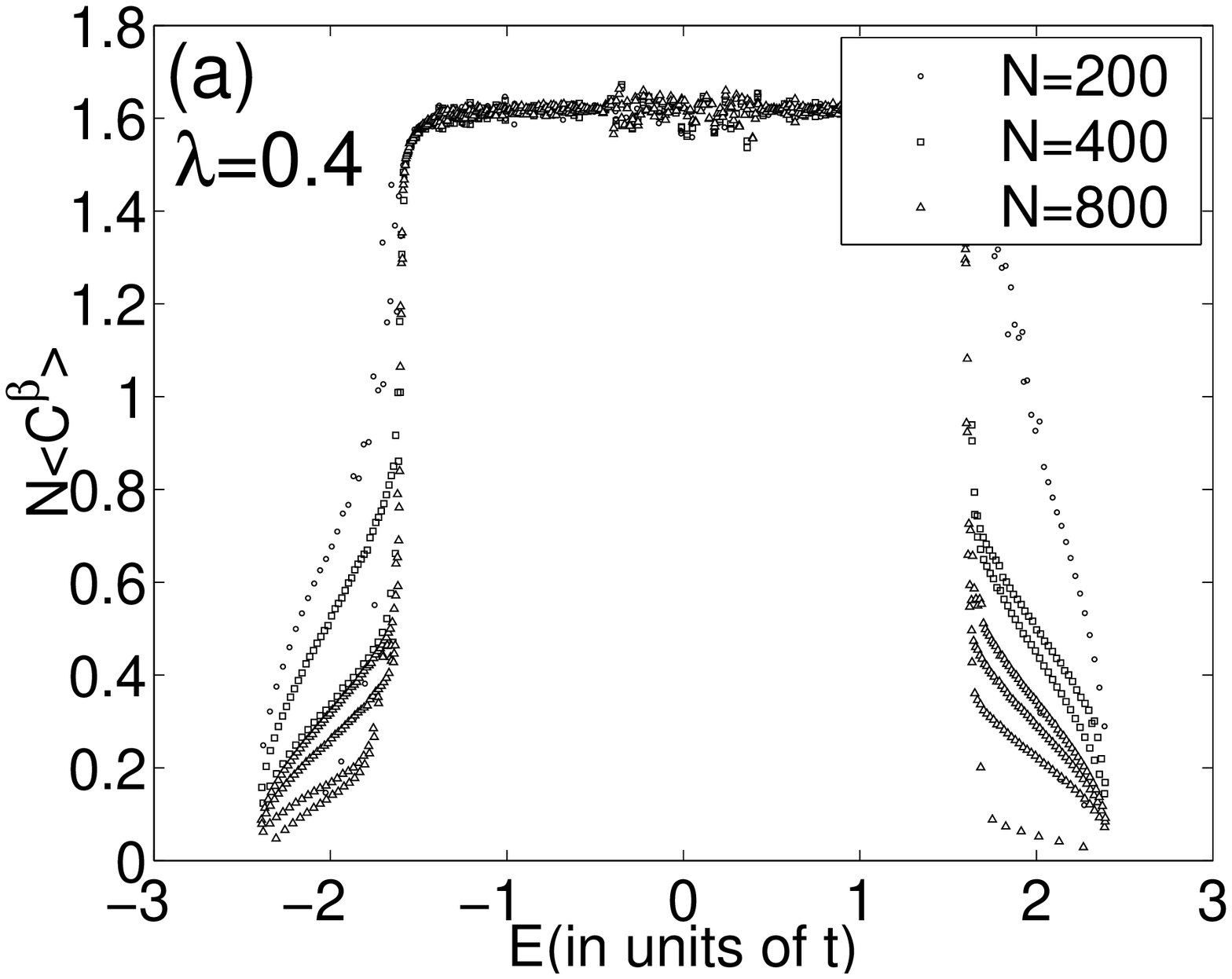}
\includegraphics[width=2.5in]{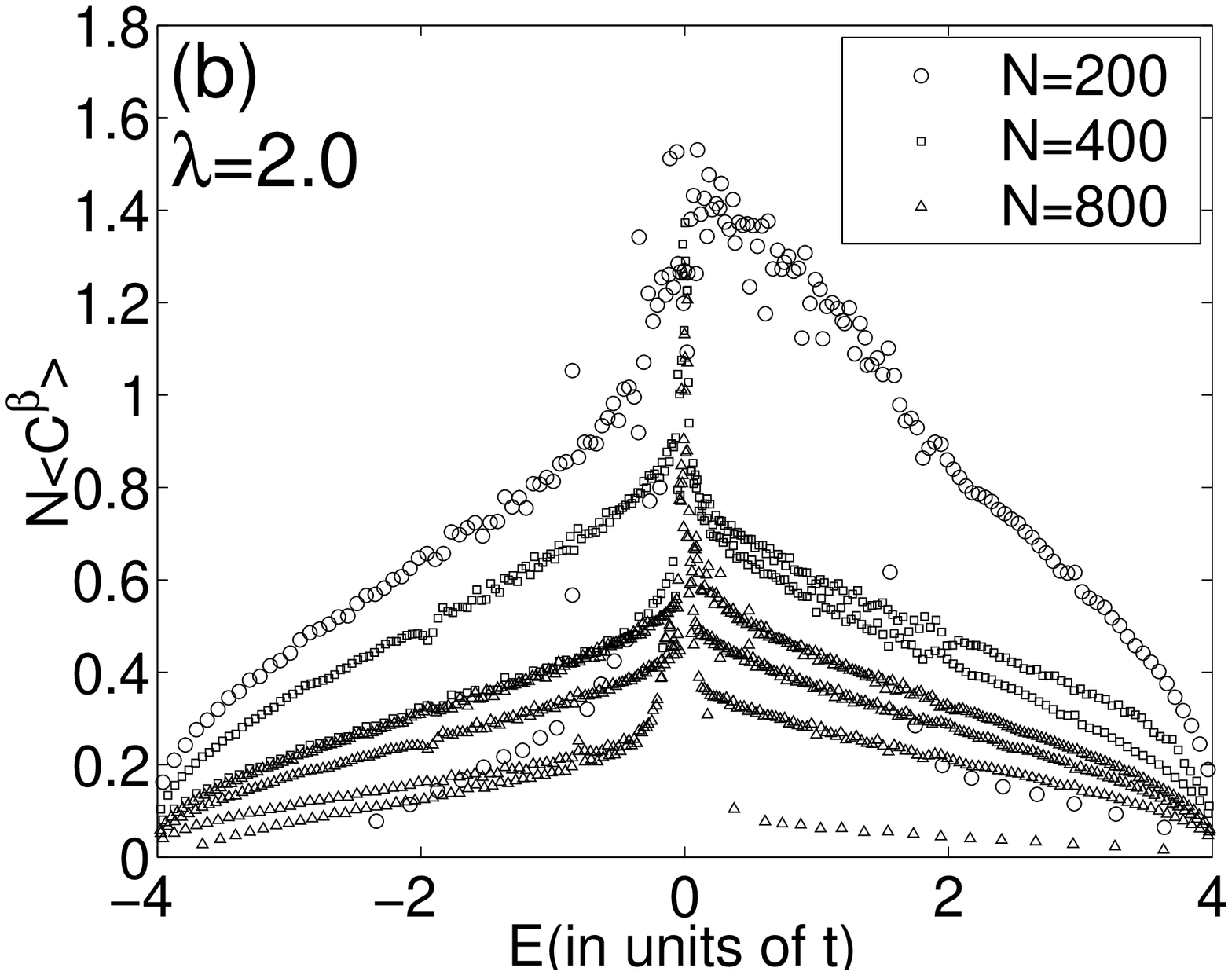}
\includegraphics[width=2.5in]{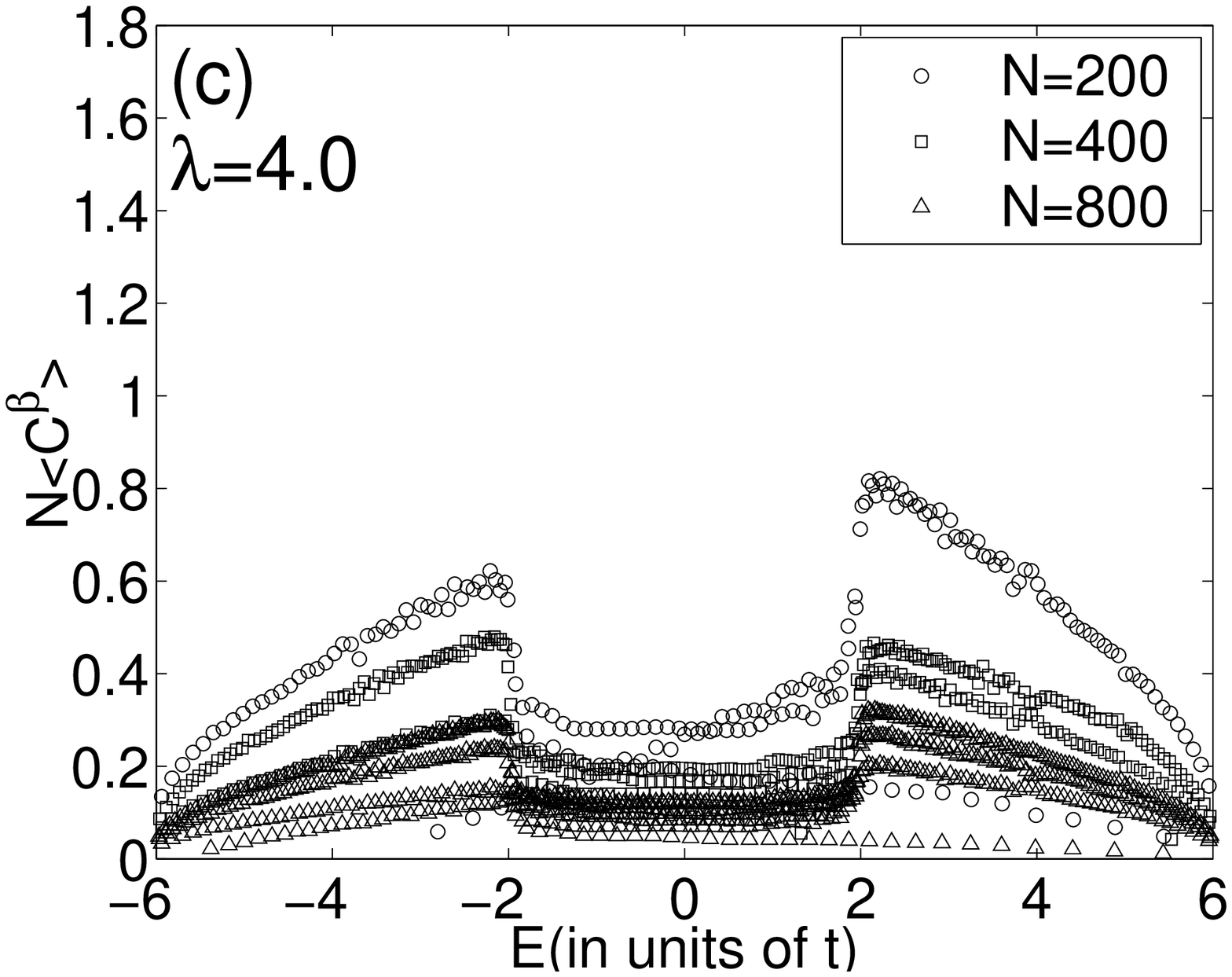}
\includegraphics[width=2.5in]{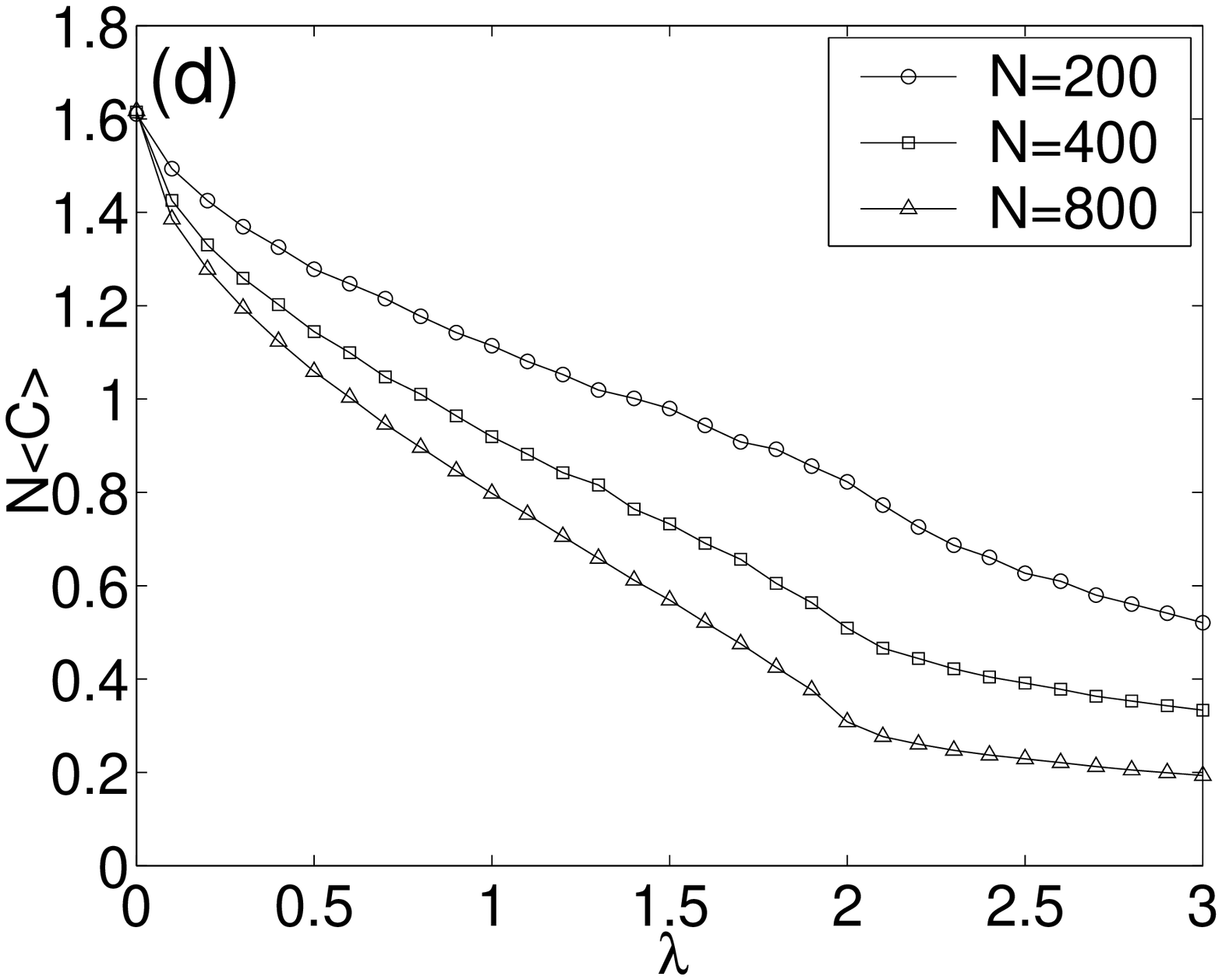}\caption{Average concurrence
$N\left\langle {C^{\beta}} \right \rangle $ of the individual
state in the slowly varying potential model as functions of
energy, with $\pi\alpha=0.2$ and $\upsilon=0.7$, (a)
$\lambda=0.4$, (b) $\lambda=2.0$ and (c) $\lambda=4.0$,
respectively. (d) Average concurrence $N\left\langle {C}
\right\rangle$ over all states varying with $\lambda$. }
\end{figure}

For $0<\upsilon <1$, it is well known that there are two mobility
edges at $E_c = \pm (2.0 - \lambda ) $ provided that
$\lambda<2.0$. It was found extended states in the middle of the
band ( $\left| E \right| < 2.0 - \lambda$ ) and localized states
at the band edge ( $2.0 - \lambda  < \left| E \right| < 2.0 +
\lambda$ ). For $\lambda>2.0$, all states are found to be
localized.

Concurrence $ N\left\langle C^\beta \right\rangle$ are plotted in
Fig.1(a), (b), (c) for $\lambda=0.4, 2.0$ and $4.0$ respectively.
From Fig.1(a), (b), it's clearly shown that there are sharp
transitions in concurrence at the mobility edges $E_c = \pm (2.0 -
\lambda )$. The transition becomes sharper when $N$ increasing. At
extended states, $N\left\langle {C^\beta}\right\rangle \approx
1.6$. At the localized states, $N\left\langle
{C^\beta}\right\rangle$ is decreasing from mobility edges to band
top (or bottom). In Fig.1 (c), all states are localized and
$N\left\langle {C^\beta}\right\rangle$ are small for all states,
which is smaller than 1.6. Comparing with Fig.1(a), (b) and (c) in
Ref.\cite{sa88}, we can found that the longer the localized length
is, the larger concurrence is.

The concurrence  $N\left\langle {C} \right\rangle$ averaged over
all states is plotted in Fig.1(d). There is a transition at
$\lambda=2.0$. The transition becomes sharper when $N$ increasing.
Obviously there is dramatically transition as $N\rightarrow\infty$
at $\lambda=2.0$.

\subsection{Random-dimer potential} 
Another interesting 1D model, which has localized and extended
states, is random-dimer model \cite{du90}. In this model, the site
energies $V_a$ and $V_b$ are assigned at random to $2n$th
($n$=integer) site with probability $q$ and $1-q$, and
$V_{2n+1}=V_{2n}$. By solving the time-depended Schr\"{o}dinger
equation, Dunlap {\textit{et al.}} \cite{du90} found that the
mean-square displacement at long times is shown to grow in time as
$t^{3/2}$ provided $-2<V_a-V_b<2$, diffusion occurs if
$V_a-V_b=\pm2$ and localization otherwise. These mean that
extended states exist when $-2<V_a-V_b<2$, and there are only
localized states when $|V_a-V_b|>2$.

Concurrence $ N\left\langle C^\beta \right\rangle$ are plotted in
Fig.2(a), (b), (c) for $V_a-V_b=1.0$, $2.0$ and $2.5$
respectively. The $V_b$ is taken as $1.0$ without loss of
generality and $q$ is equal to $0.5$ corresponding to most random
situation. The results of Fig.2 are obtained for average of 200
samples \cite{gl04}. The averages with more samples give same
results. There are two bumps in concurrence when $V_a-V_b=1.0$.
There is no obvious mobility edge in this model, so there is no
sharp transition in concurrence as that in the slowly varying
potential model. There are two jumps in concurrence when
$V_a-V_b=2.0$. For $V_a-V_b=2.5$, concurrences are small for all
states.

Average concurrence $ N\left\langle C \right\rangle$ as functions
of $V_a-V_b$ is plotted in Fig.2(d). There is a jump in the
concurrence at $V_a-V_b=2.0$, which is in accordance with the
critical value of $V_a-V_b$ obtained by dynamical method.

\begin{figure}
\includegraphics[width=2.5in]{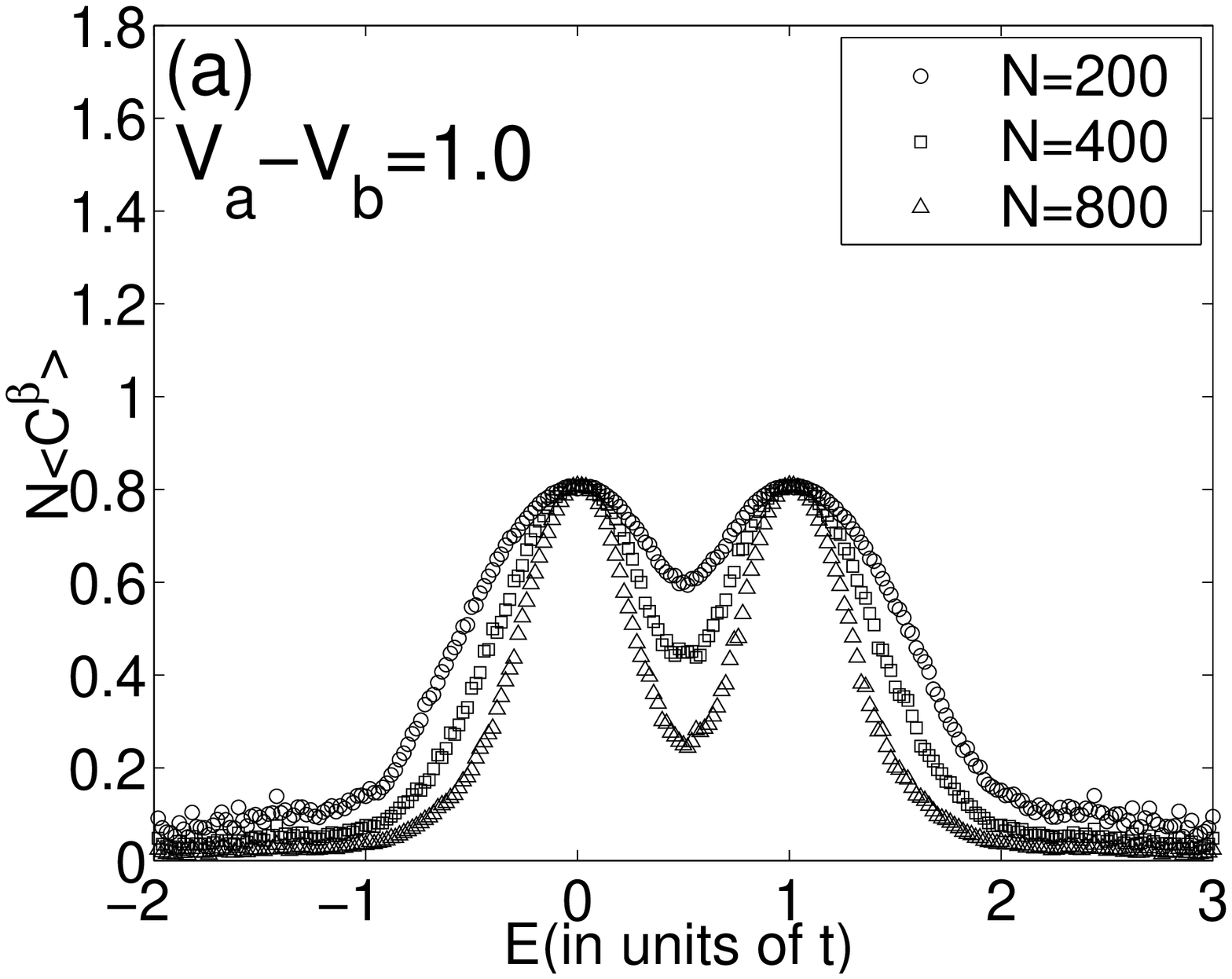}
\includegraphics[width=2.5in]{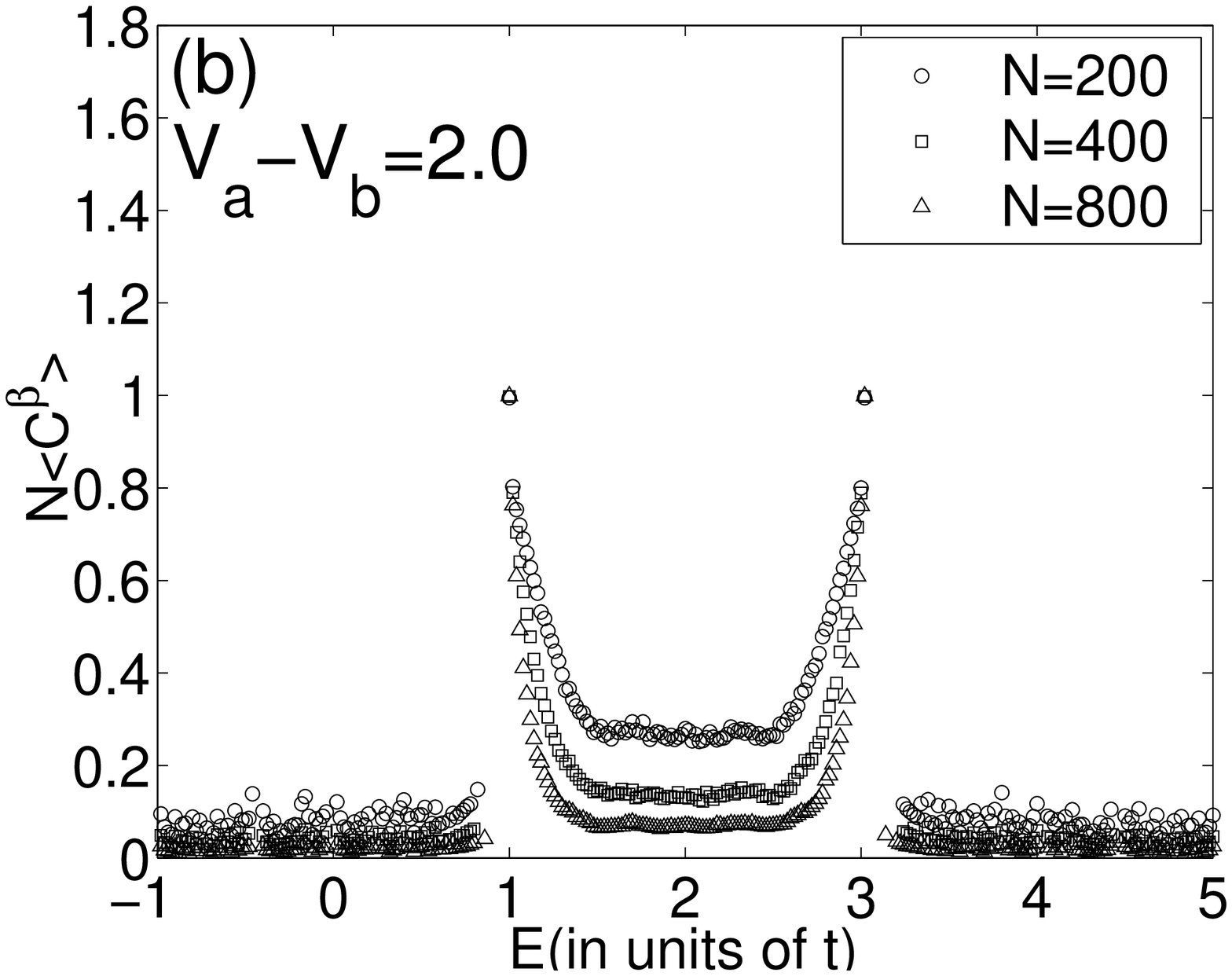}
\includegraphics[width=2.5in]{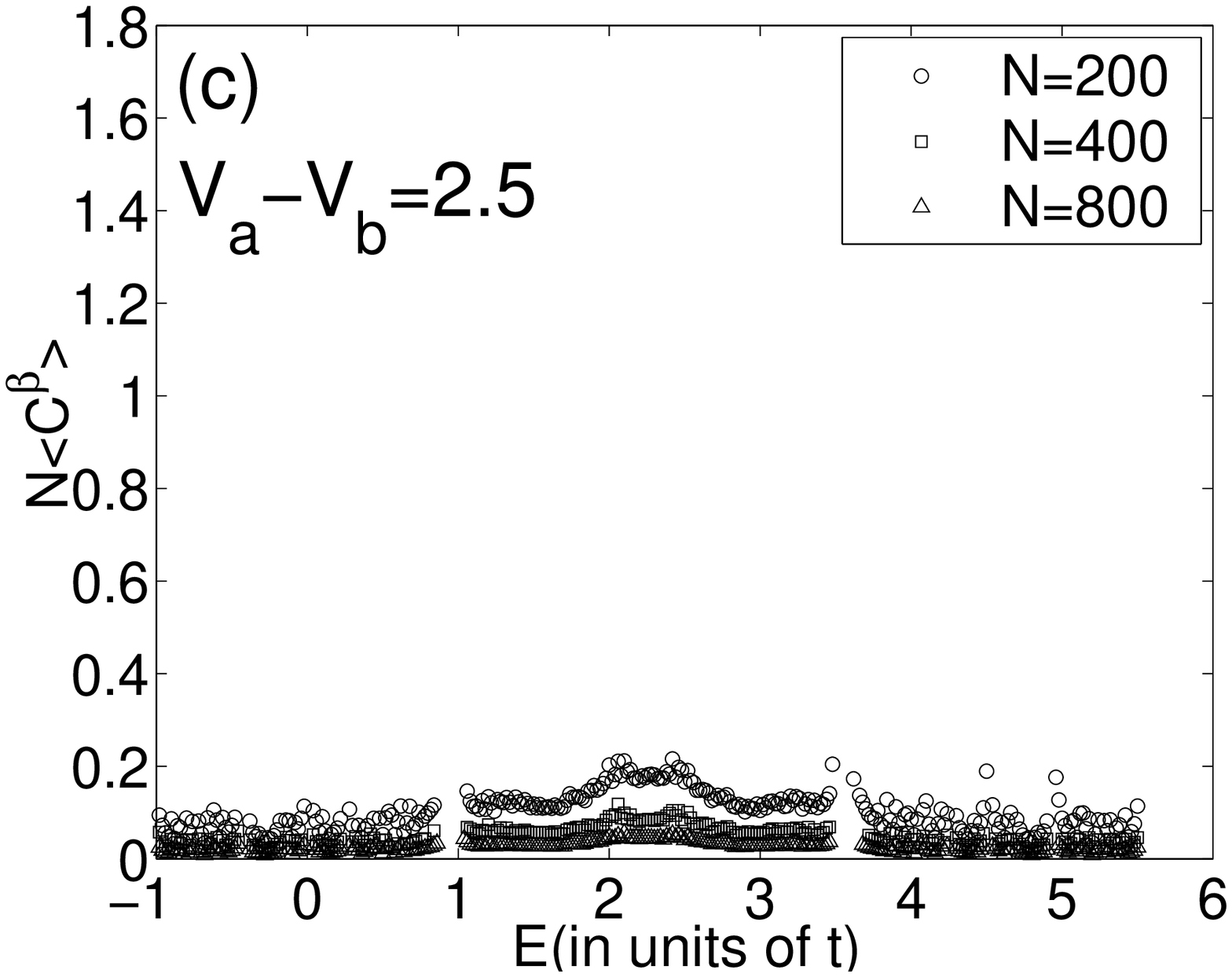}
\includegraphics[width=2.5in]{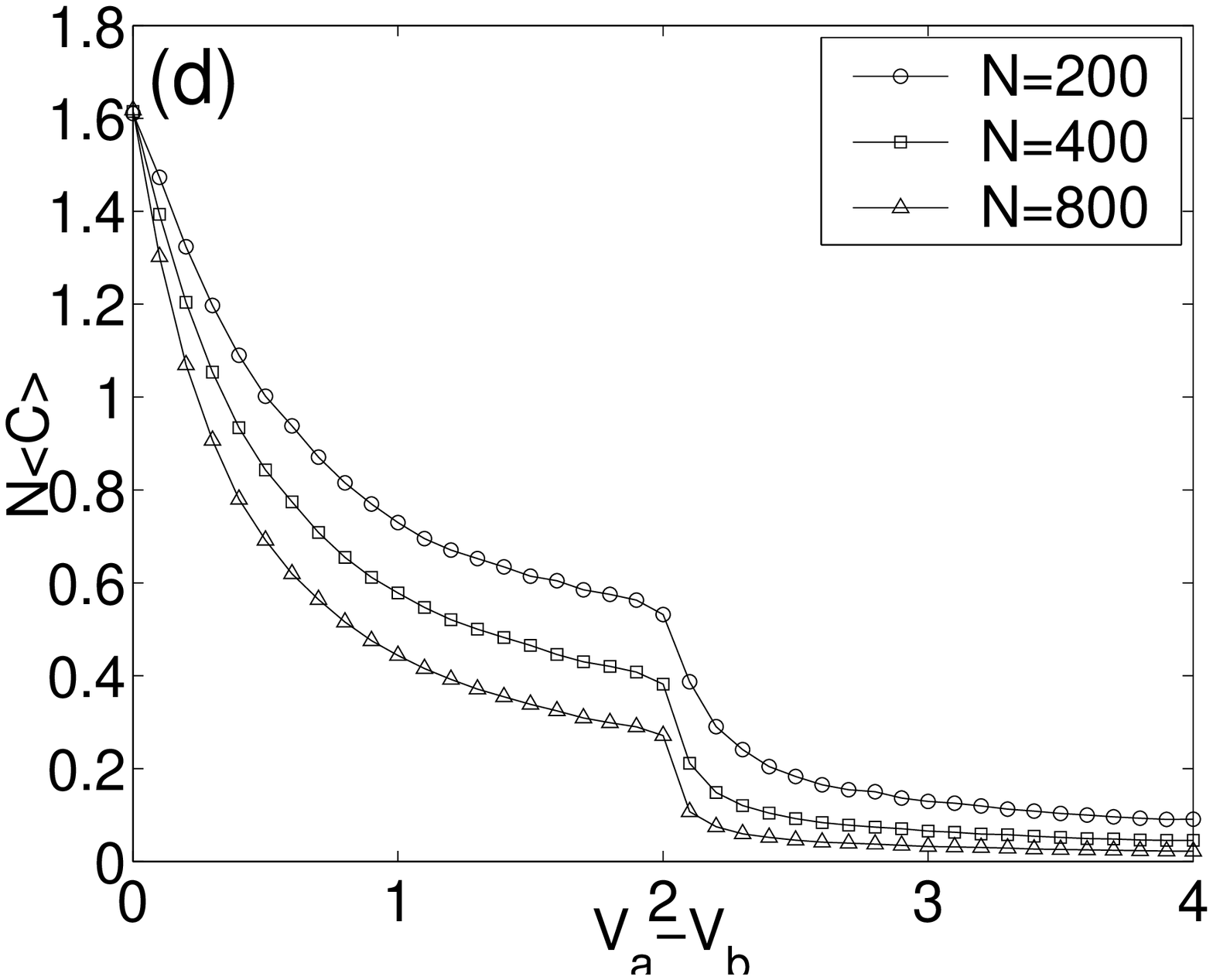}\caption{Average concurrence
$N\left\langle {C^{\beta}} \right\rangle $ of the individual state
in the random-dimer model as functions of energy, (a)
$V_a-V_b=1.0$, (b) $V_a-V_b=2.0$ and (c) $V_a-V_b=2.5$,
respectively. (d) Average concurrence $N\left\langle {C}
\right\rangle$ over all states in the dimer model as functions of
$V_a-V_b$.}
\end{figure}

\subsection{Long-range correlated disordered potential } 

Recently, another kinds of disordered potential is studied
extensively. The potential is self-affine Gaussian potential with
Hurst exponent $0<H<1$, such that $
 < (V_m  - V_n )^2  >  = \Delta ^2 \left| {m - n} \right|^{2H}$.
Such sequence of potential can be generated by fractional Brownian
motion. To generate the trace of a fractional Brownian motion, an
approach based on the use of discrete Fourier transforms to
construct such long-range correlated sequences can be applied. The
on-site energies can be given by the relation\cite{mo98}:
\begin{equation}
V_i  = \sum\limits_{k = 1}^{N/2} {[k^{ - \alpha } \left|
{\frac{{2\pi }}{N}} \right|^{(1 - \alpha )} ]} ^{1/2} \cos
(\frac{{2\pi ik}}{N} + \varphi _k ),
\end{equation}                
where $N$ is the number of sites and $\varphi_k$  are $N/2$
independent random phases uniformly distributed in the interval
$[0,2\pi ]$. The sequence usually has an approximate power-law
spectral density of the form $S(k) \propto 1/k^\alpha$, where
$S(k)$ is the Fourier transform of the two-point correlation $ <
V_i V_j  > $. Here $\alpha=2H+1$. Same as in Ref. \cite{mo98}, we
normalized the energy sequence to have $<V_n>=0$ and $\Delta V =
\sqrt {\left\langle {V_n^2 } \right\rangle  - \left\langle {V_n }
\right\rangle ^2 }  = 1 $.

It is shown \cite{mo98} that when $\alpha<2.0$, all states are
localized; when $\alpha>2.0$, localized states occur in the edges
of the band and the extended states in the middle of the band,
separated by mobility edge.

The relations between concurrence $ N\left\langle C^\beta
\right\rangle$ and eigenenergy $E$ are plotted in Fig.3(a), (b)
and (c) for $\alpha=5.0, 2.0$ and $1.0$ respectively. Here 200
samples of random $\varphi_k$ are averaged. For $\alpha=5.0$,
$N\left\langle C^\beta\right\rangle\approx1.6$ in the middle
states and $N\left\langle C^\beta\right\rangle$ is gradually
decreasing from center to band top (or band bottom). The
eigenenergies region of extended states is decreasing as $\alpha$
becomes smaller \cite{mo98}. When $\alpha=2.0$, the concurrence is
near $1.6$ only at band center, which is shown clearly in inset of
Fig.3(b). When $\alpha=1.0$, concurrence for all states is small,
which corresponds to localized states.

Concurrence averaged over all states is plotted in Fig.3(d). There
is an inflexion when $\alpha$ is near $2.0$ in the figure. Same as
in model with slowly varying potential, the bigger $N$ is, the
transition is sharper.

\begin{figure}
\includegraphics[width=2.5in]{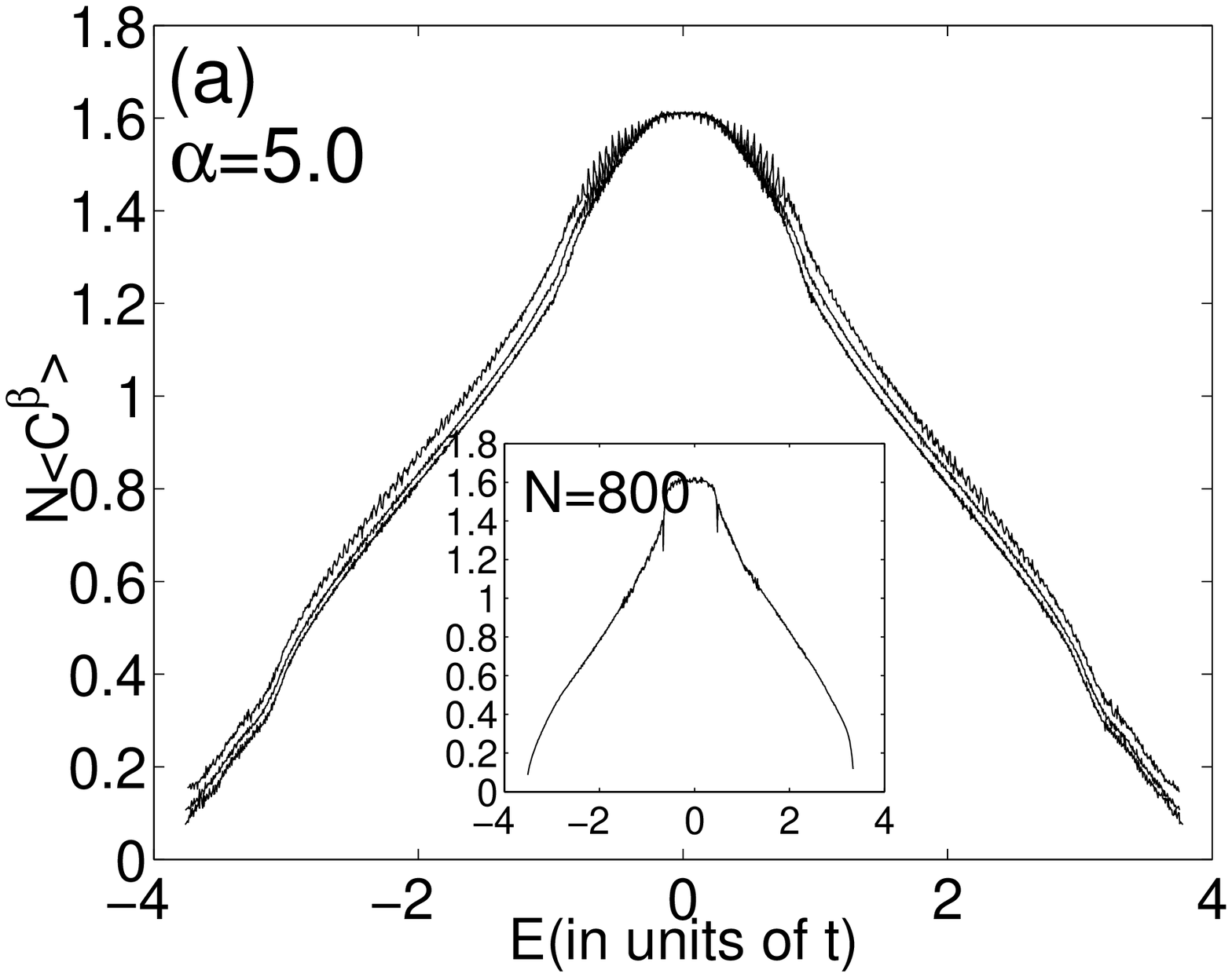}
\includegraphics[width=2.5in]{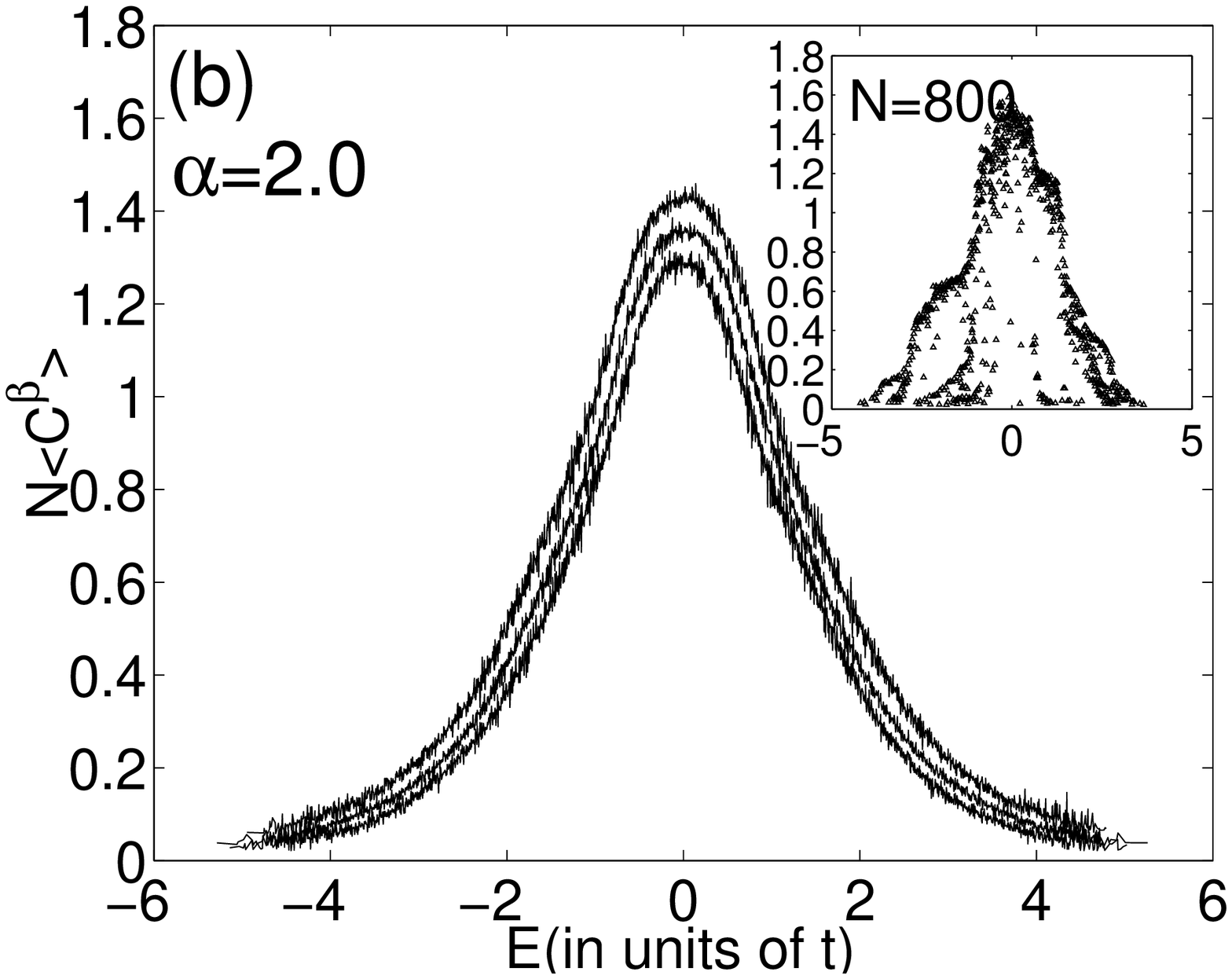}
\includegraphics[width=2.5in]{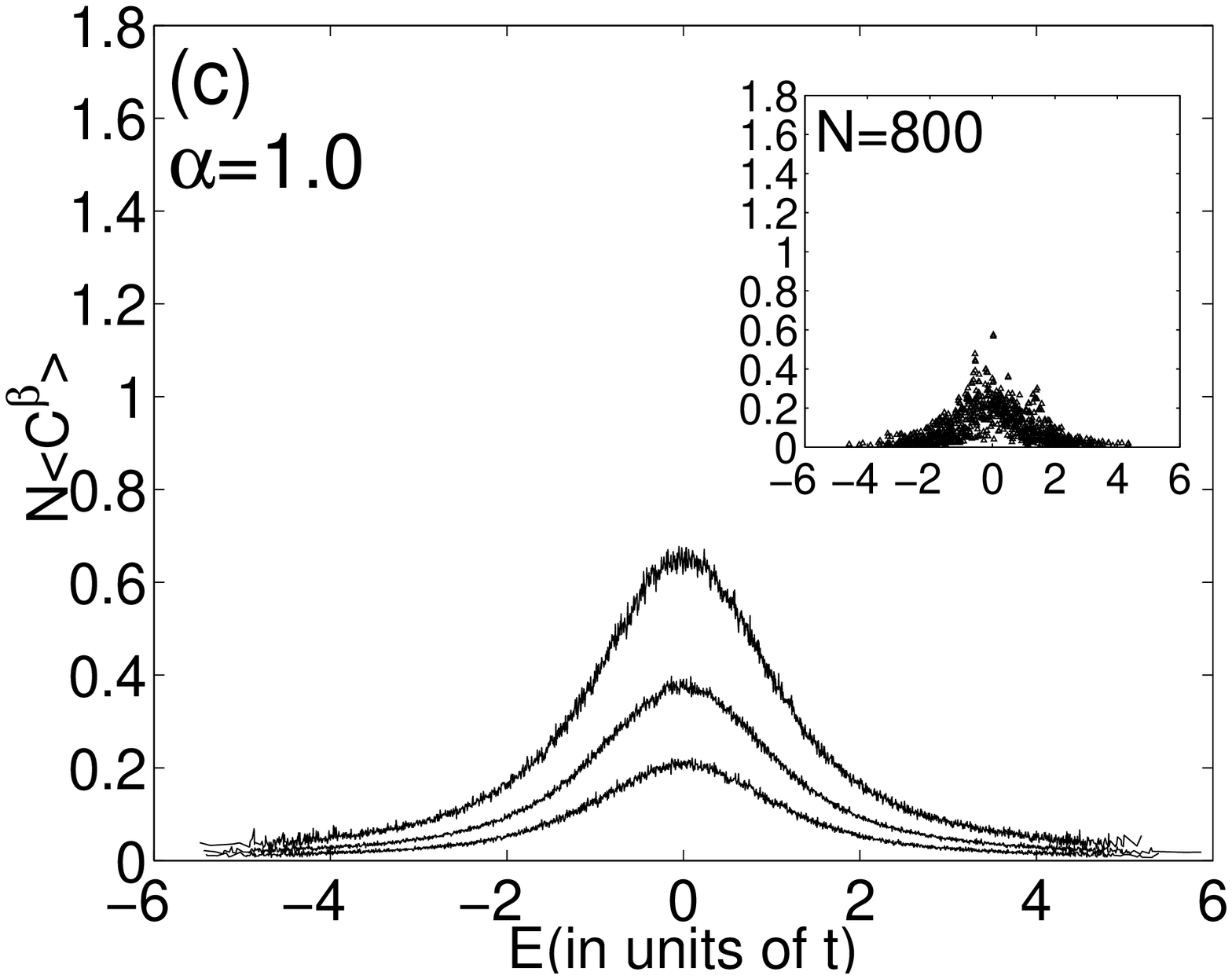}
\includegraphics[width=2.4in]{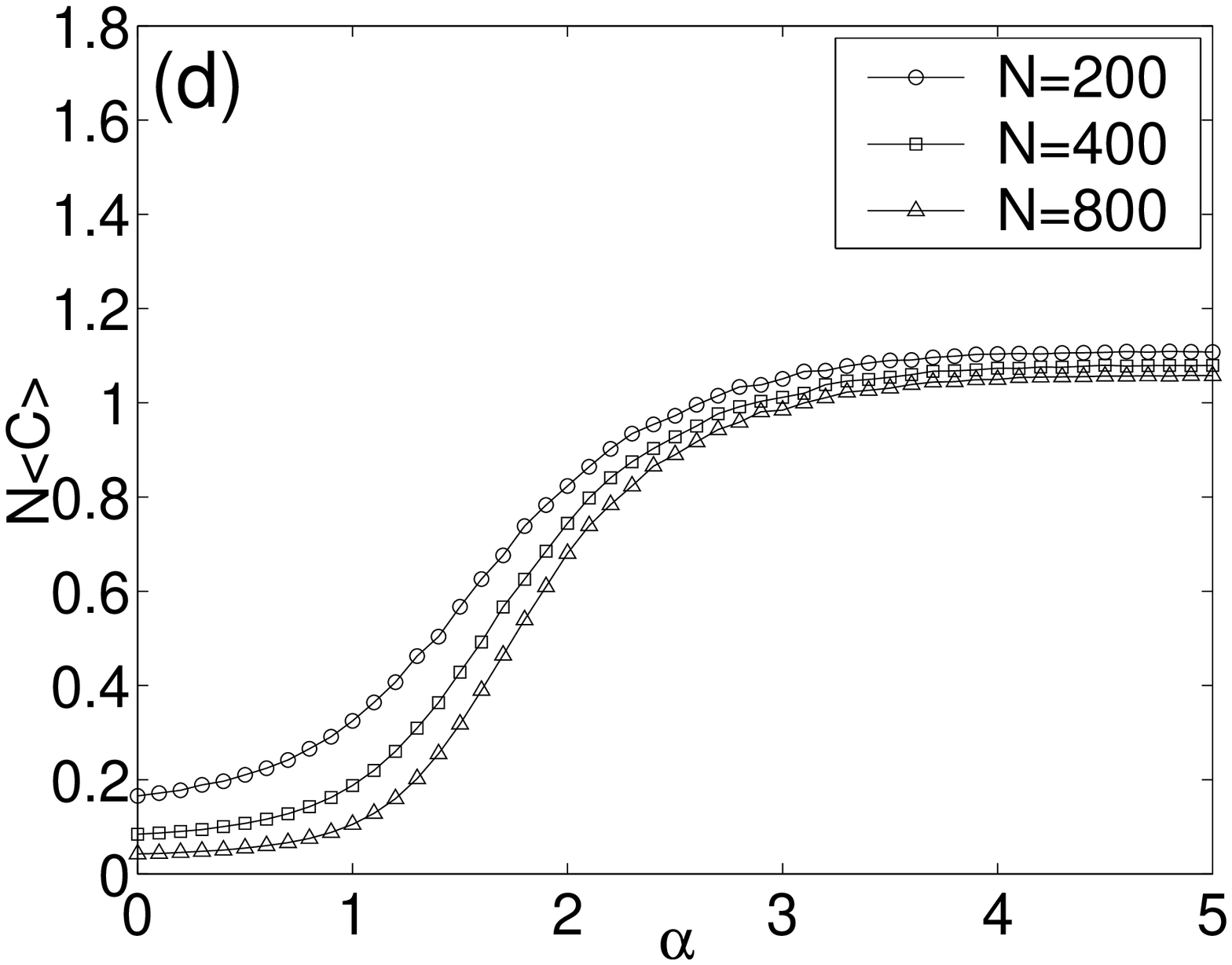}\caption{Average concurrence
$N\left\langle {C^{\beta}} \right\rangle$ of the individual state
in the long-range correlated disordered potential model as
functions of energy for (a) $\alpha=5.0$, (b) $\alpha=2.0$ and (c)
$\alpha=1.0$, respectively. The insets show concurrence for a
typical disorder realization at N=800. (d) Average concurrence
$N\left\langle {C} \right\rangle$ over all states as functions of
$\alpha$. In all figures, from up to low, $N=200$, $400$ and
$800$. }
\end{figure}

\subsection{Random potential with long-range hopping}

The random potential with long-range hopping has received
considerable attention recently. The Hamiltonian of such 1D
tight-binding model is expressed as
\begin{equation}
H = \sum\limits_n {\varepsilon _n } \left| n \right\rangle
\left\langle n \right| + \sum\limits_{n \ne m} {J_{mn} } \left| n
\right\rangle \left\langle m \right|,
\end{equation} 
where $\varepsilon_n$ is energy level at \emph{n}th site ,
uniformly distributed in interval $[-W/2,W/2]$ and $J_{mn}  =
J/\left| {m - n} \right|^\mu $ ($J_{mm}\equiv0$) is the long-range
hopping amplitude. We will adopt $J$ as energy units without loss
of generality. The periodic boundary condition is applied.

By using a supersymmetric method combined with a renormalization
group analysis, Rodriguez \emph{et al.} \cite{ro03} have shown the
existence of extended states for energies within a range near the
band top in one and two dimensional Anderson models. They found
that MIT occurs only within the range of $d<\mu<3d/2$ in the
thermodynamic limit, no matter how large the value of $W$ is, here
$d$ is the geometric dimensionality of the system. Using finite
size scaling analysis combined with the transfer matrix method,
Xiong and Zhang \cite{xi03}found that there exists MIT at critical
value $\mu_c$ for some $W$.

The concurrence $ N\left\langle C^\beta\right\rangle$ as functions
of energy are plotted in Fig.4 (a), (b) and (c) for $\mu=1.1$,
$1.5$ and $1.7$ respectively. Here $W=5$ is taken as example and
the phenomena are similar for other $W$. The results are obtained
for average of $200, 100, 50$ samples \cite{gl04} for $N=200, 400,
800$ respectively. When $\mu=1.1$ and $1.5$, concurrence $
N\left\langle C^\beta\right\rangle$ is large for states near the
band top, which means there exist extended states. It is small for
the states near the band bottom. This is quite different from that
of above three models. For $\mu=1.7$, the concurrence is small for
all states, which means all states are localized.

The concurrence averaged over all states is plotted in Fig.4(d).
From the inset, we can get the inflexion for $\mu$ near $1.70$,
which is consistence with the upper limit for critical value
$\mu_c$ obtained in Ref.\cite{xi03}.

\begin{figure}
\includegraphics[width=2.5in]{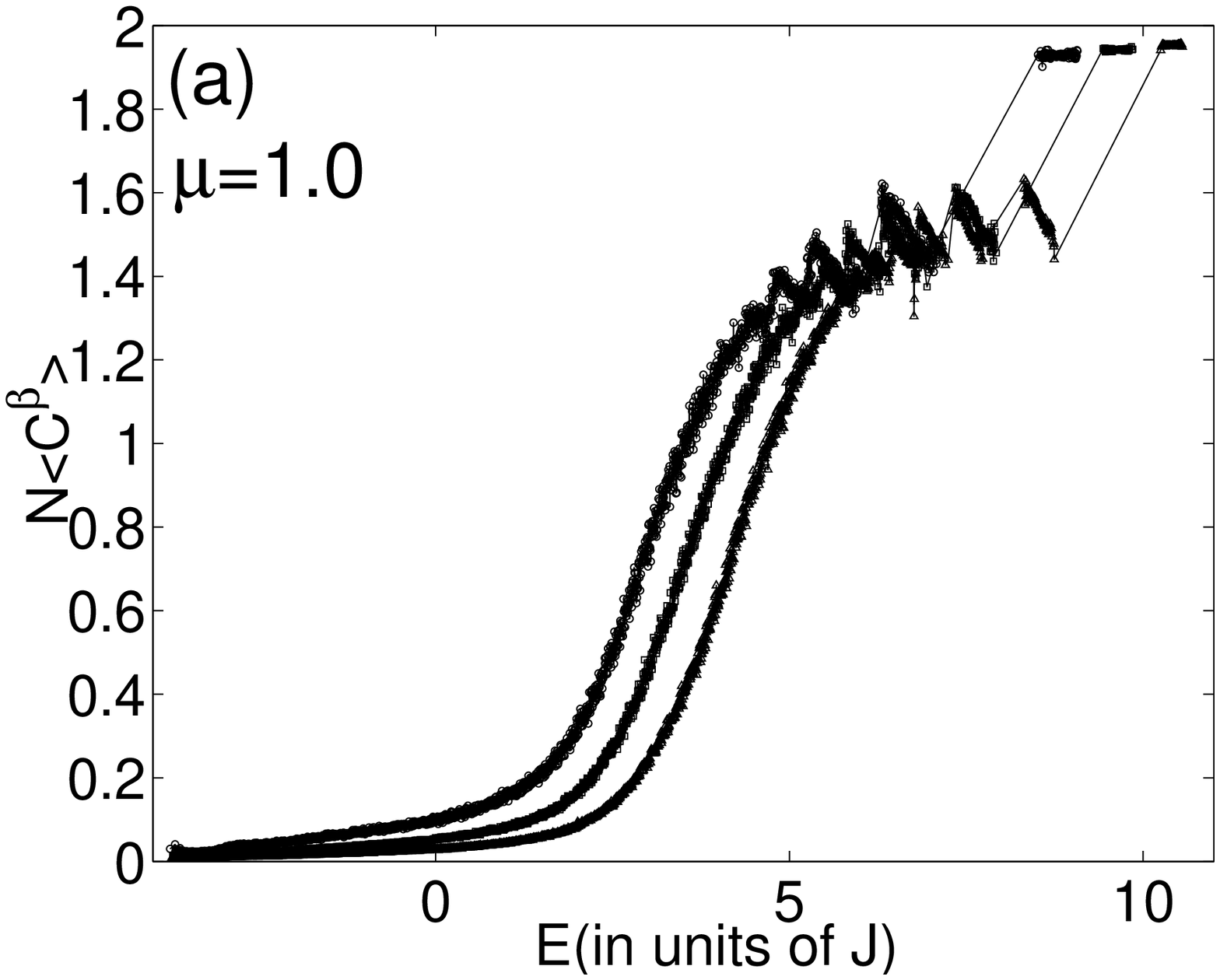}
\includegraphics[width=2.5in]{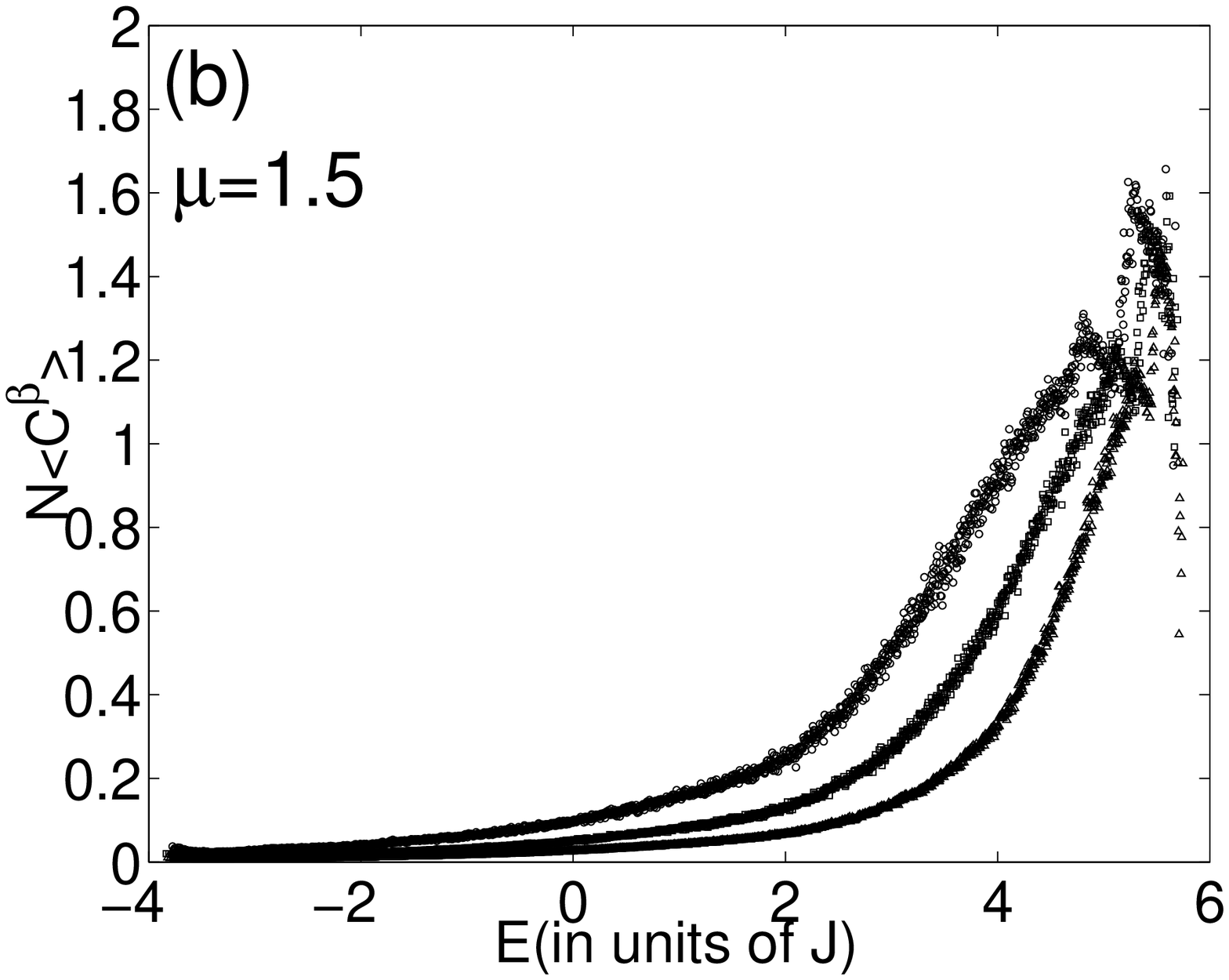}
\includegraphics[width=2.5in]{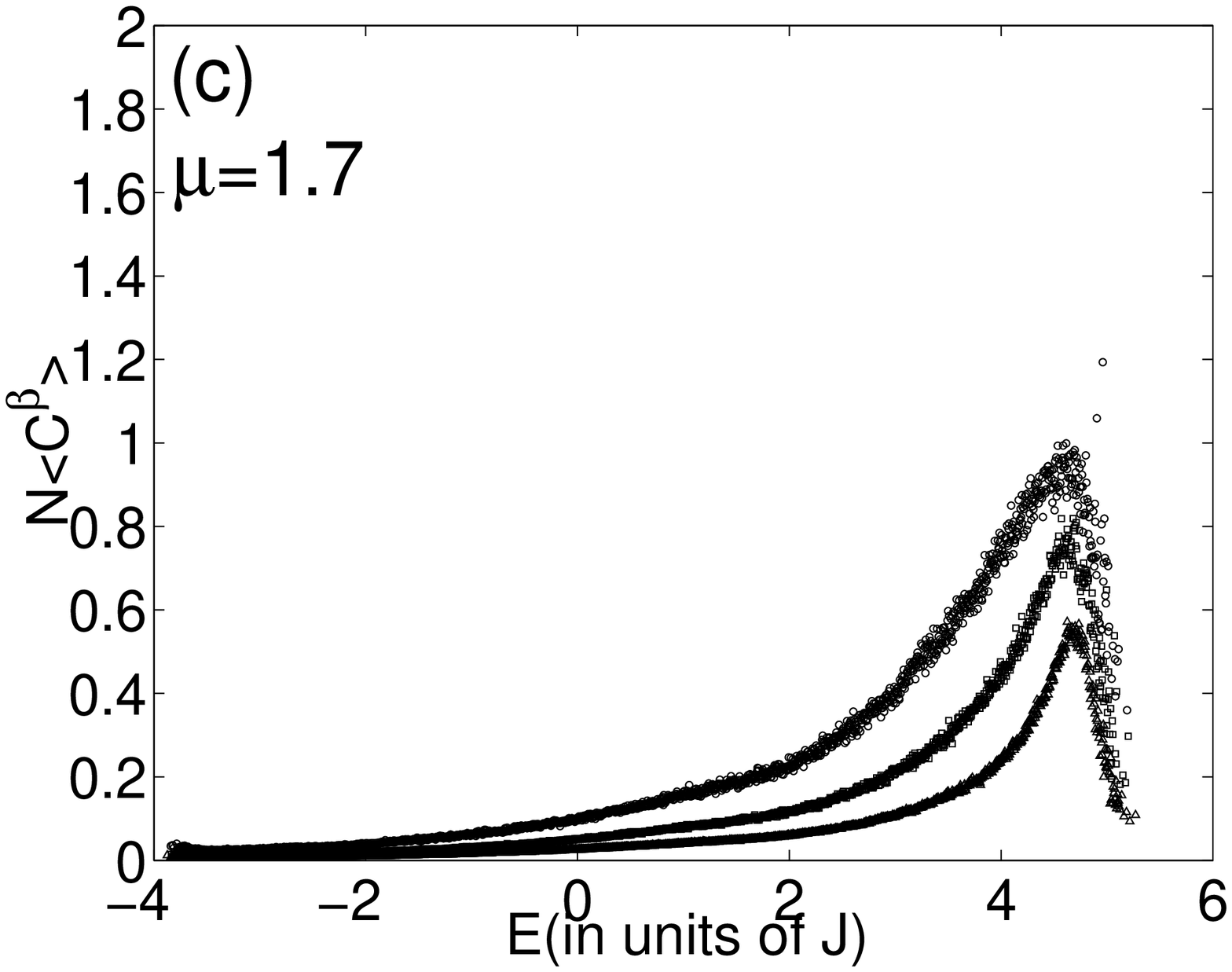}
\includegraphics[width=2.5in]{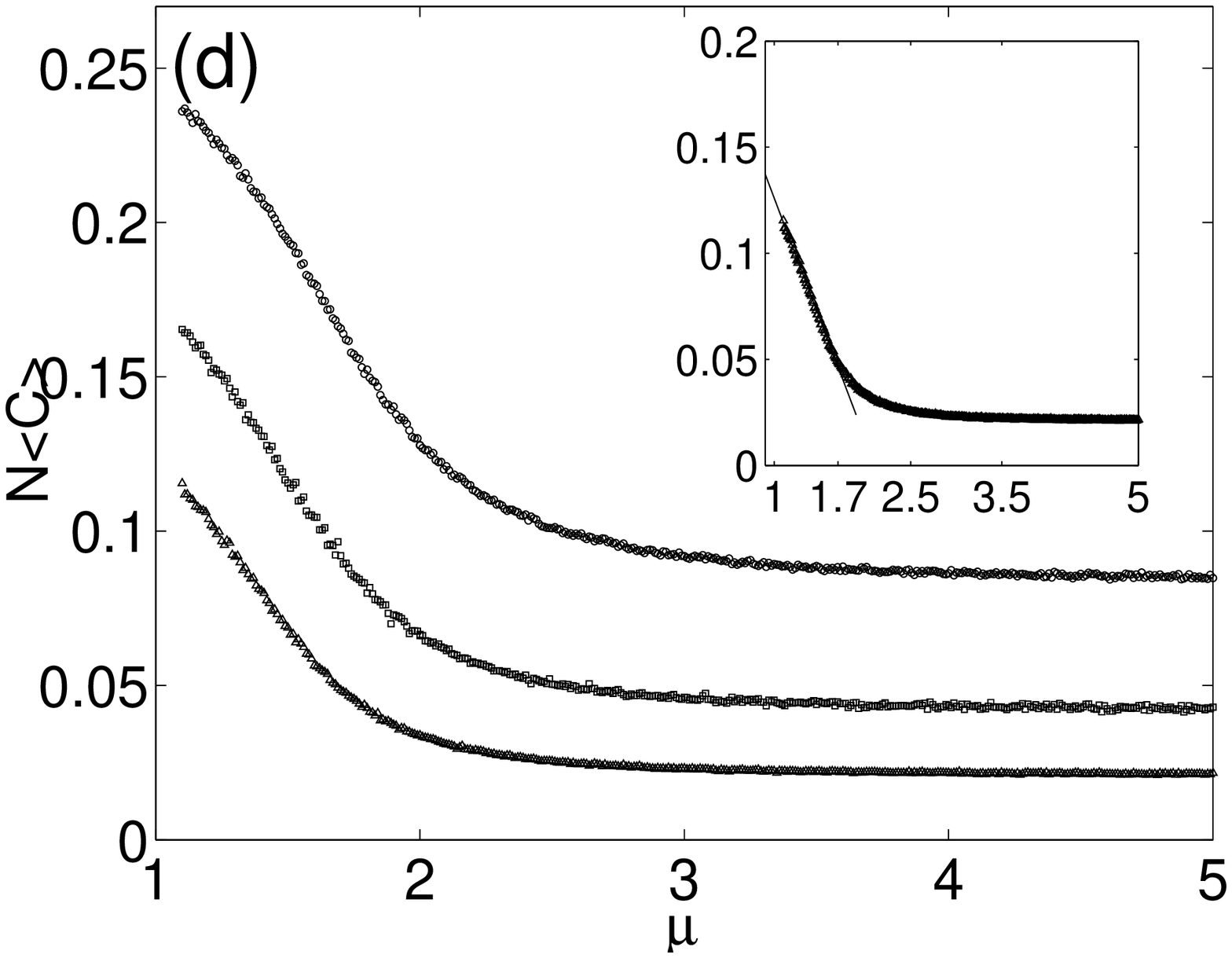}
\caption{Average concurrence $N\left\langle {C^{\beta}}
\right\rangle$ of the individual state in the model with
long-range hopping as functions of energy, with  $W=5.0$, (a)
$\mu$=1.1, (b) $\mu=1.5$ and (c) $\mu=1.7$, respectively.
(d)Average concurrence $N\left\langle {C} \right\rangle$ over all
states as functions of $\mu$. The inset shows concurrence when
$N=800$. The line is linear fitted line for $\mu<1.5$ . In all
figures, from up to low, $N=200, 400, 800$.}
\end{figure}

\section{\label{sec4}Conclusion}
 Using the measure of concurrence,
 mode entanglement sharing in one-particle states
 in four  kinds of models is studied numerically.
Concurrence $N\left\langle C^\beta\right\rangle$ at a given state
and $ N\left\langle C\right\rangle$
 averaged over all states are investigated. For $N\left\langle C^\beta\right\rangle$,
 the concurrence is large in extended states and small in localized states. There is a sharp
 transition in the concurrence at mobility edge. $N\left\langle C^\beta\right\rangle$ gives the information about the
 localization behavior of the given eigenstate $\beta$.  From the curves of
 the $N\left\langle C\right\rangle$ vs the parameter of the models, which is
 $\lambda$,  $V_a-V_b$, $\alpha$, and $\mu$ for slowly varying potential,
 random-dimer potential, long-range correlated
 disordered potential, and long-range hopping random potential,
 respectively, we can found clearly that there is an inflexion (or jump) at a critical parameter
 value, which is in accordance with that obtained by other
 methods. When parameter value is greater (smaller) than the critical parameter value,
 the system has only localized eigenstates, while when
parameter value is smaller (greater) than the critical parameter
value, the system has both localized and delocalized states, which
is different from that of one-dimensional Harper model. The
inflexion or transition point in the curve of $N\left\langle
C\right\rangle$ versus parameter of systems corresponds to the
disappear of delocalized states. Therefore mode entanglement can
be a new index to reflect MIT.

\begin{acknowledgments}
We would like to thank the referee for helpful suggestions and
comments. This work is partly supported by the National Nature
Science Foundation of China under Grant Nos. 90203009 and
10175035, by the Nature Science Foundation of Jiangsu Province of
China under Grant No. BK2001107, and by the Excellent Young
Teacher Program of MOE, P.R. China.
\end{acknowledgments}


\begin{thebibliography}{99}
\bibitem{gl04}For random potentials, we use the following
definition of concurrence:$\langle {C^\beta}
\rangle=\overline{\frac{1}{d}\sum\limits_{i < j} {C_{ij}^\beta
}}=\frac{1}{K}\frac{1}{d}\sum\limits_{k=1}^K\sum\limits_{i < j}
{(C_{ij}^\beta)_k }$, here $\overline{X}$ is denoted as random
average, $K$ is the number samples, and $\beta$ denotes a given
state.
\end{thebibliography}
\end{document}